\documentclass[twocolumn]{aastex631}

\hypersetup{linkcolor=red,citecolor=blue,filecolor=cyan,urlcolor=blue}

\received{{\today}}

\usepackage{graphicx}
\usepackage{ragged2e}
\usepackage{mathtools}

\providecommand{\e}[1]{\ensuremath{\times 10^{#1}}}

\submitjournal{ApJ}

\graphicspath{{./}{figures/}}

\shortauthors{Salman et al.}

\begin{document}

\title{A Survey of Coronal Mass Ejections Measured In Situ by Parker Solar Probe During 2018--2022}

\correspondingauthor{Tarik M. Salman}
\email{tsalman@gmu.edu}

\author[0000-0001-6813-5671]{Tarik Mohammad Salman}
\affiliation{Department of Physics and Astronomy, George Mason University, 4400 University Dr., MSN 3F3, Fairfax, VA 22030, USA}
\affiliation{Heliophysics Science Division, NASA Goddard Space Flight Center, Greenbelt, MD 20771, USA}

\author[0000-0003-0565-4890]{Teresa Nieves-Chinchilla}
\affiliation{Heliophysics Science Division, NASA Goddard Space Flight Center, Greenbelt, MD 20771, USA}

\author[0000-0002-6849-5527]{Lan K. Jian}
\affiliation{Heliophysics Science Division, NASA Goddard Space Flight Center, Greenbelt, MD 20771, USA}

\author[0000-0002-1890-6156]{No\'{e} Lugaz}
\affiliation{Institute for the Study of Earth, Oceans, and Space, University of New Hampshire, Durham, NH 03824, USA}

\author[0000-0003-1758-6194]{Fernando Carcaboso}
\affiliation{Postdoctoral Program Fellow, NASA Goddard Space Flight Center, Greenbelt, MD 20771, USA}

\author[0000-0001-9992-8471]{Emma E. Davies}
\affiliation{Austrian Space Weather Office, GeoSphere Austria, Graz 8020, Austria}

\author[0000-0003-3396-196X]{Yaireska M. Collado-Vega}
\affiliation{Heliophysics Science Division, NASA Goddard Space Flight Center, Greenbelt, MD 20771, USA}

\begin{abstract}

\justify
We present a statistical investigation of the radial evolution of 28 interplanetary coronal mass ejections (ICMEs), measured in situ by the Parker Solar Probe (PSP) spacecraft from 2018 October to 2022 August. First, by analyzing the radial distribution of ICME classification based on magnetic hodograms, we find that coherent configurations are more likely to be observed close to the Sun. In contrast, more complex configurations are observed farther out. We also notice that the post-ICME magnetic field is more impacted following an ICME passage at larger heliocentric distances. Second, with a multi-linear robust regression, we derive a slower magnetic ejecta (ME) expansion rate within 1~au compared to previous statistical estimates. Then, investigating the magnetic field fluctuations within ICME sheaths, we see that these fluctuations are strongly coupled to the relative magnetic field strength gradient from the upstream solar wind to the ME. Third, we identify ME expansion as an important factor in forming sheaths. Finally, we determine the distortion parameter (DiP) which is a measure of magnetic field asymmetry in an ME. We discover lower overall asymmetries within MEs. We reveal that even for expanding MEs, the time duration over which an ME is sampled does not correlate with DiP values, indicating that the aging effect is not the sole contributor to the observed ME asymmetries.  

\end{abstract}


\section{Introduction} \label{sec:intro}


\justify
Investigating the evolution of interplanetary counterparts of coronal mass ejections (CMEs), hereafter ICMEs, is crucial to develop an improved generalization of the embedded complexities in ICME propagation. The dynamic propagation of ICMEs in the heliosphere is governed by its nature of expansion and interactions with the background solar wind and other transients therein \citep[see][for a review]{manchester2017}. However, the degree of their influence on ICME evolution is also reliant upon the heliospheric conditions that the ICME encounters during propagation \citep[e.g.,][]{temmer2011,kilpua2012,liu2013}. The complex manifestations of ICME propagation in the structured solar wind, such as deflections from an expected radial path \citep[e.g.,][]{wang2004,rodriguez2011,lugaz2012,isavnin2014,wang2014}, deformations and distortions \citep[e.g.,][]{riley2004,temmer2014,owens2017,vrsnak2019,owens2020,davies2021}, erosion \citep[e.g.,][]{dasso2006,lavraud2014,ruffenach2015,pal2020,pal2021,stamkos2023}, resulting from the co-action of internal and external factors complicate the conceptualization and quantification of ICME evolution in the inner heliosphere.

\justify
Our knowledge of the governing physics of CME/ICME propagation has been primarily achieved through remote-sensing observations closer to the Sun and in situ measurements at 1~au. Much fruitful science has been gleaned from in situ measurements of ICMEs. Statistical investigations of ICMEs, based on single-point measurements at different heliocentric distances have provided us with significant information about average ICME characteristics in the inner heliosphere with robust data sets \citep[e.g.,][]{jian2006,richardson2010,jian2011,vech2015,winslow2015,good2016,jian2018,nieveschinchilla2018,nieveschinchilla2019}. In addition, reconstruction techniques and theoretical models \citep[e.g.,][]{burlaga1988,lepping1990,hu2002,demoulin2008,nieveschinchilla2016} and superposed epoch analysis \citep[e.g.,][]{masias2016,rodriguez2016,jian2018,janvier2019,carcaboso2020,lanabere2020,regnault2020,salman2020b} have been performed to deduce the global ICME properties by using a single spacecraft. However, such measurements provide a limited and localized view of the large and complex ICME structure, covering several tenths of an au in the radial direction near/at 1~au \citep[e.g.,][]{jian2006,richardson2010,jian2018}. Thus, a single 1D trajectory through an ICME is not enough to determine its global 3D structure \citep[see][]{alhaddad2013}. Also, the aforementioned approaches are based on assumptions of ICME geometries and spacecraft trajectories. They are not always able to provide information regarding the plethora of physical processes that the ICME might have undergone during its propagation \citep[e.g.,][]{reinard2012,lugaz2020a}.

\justify
To better understand the global configuration and evolution of ICMEs, multi-spacecraft measurements with various radial and longitudinal separations are crucial \citep[][]{winslow2022,scolini2023}. Statistical studies of in situ measurements of distinct ICMEs \citep[e.g.,][]{bothmer1998,liu2005,leitner2007,jian2008,jian2008b,winslow2015,janvier2019} or the same ICME \citep[e.g.,][]{kilpua2011,good2019,vrsnak2019,lugaz2020a,salman2020a,davies2022,scolini2022,regnault2023} from multiple spacecraft at different heliocentric distances have provided important information and a global perspective of the evolution of ICME structures in the inner heliosphere. 

\justify
It is important to note that multi-spacecraft measurements of the same ICME are still difficult to attain due to the limited number of space assets and lack of continuous solar wind measurements at different locations. With the appropriate alignment of existing spacecraft in space, it is possible to track the propagation of a CME from the Sun to Mercury, 1~au orbit, Mars, Saturn, New Horizons (at 32~au), and Voyager 2 (at 110~au), as shown in \citet{witasse2017}. It has also been seen that it is even possible to encounter the same ICME at points in space that have a 110$^{\circ}$ longitudinal separation \citep[see][]{carcaboso2024}. However, such multi-spacecraft encounters also depend on fortuitous radial and longitudinal lineups of space probes. Also, factors such as inclination \citep[e.g.,][]{kilpua2009,davies2021}, longitudinal separation \citep[e.g.,][]{farrugia2011,kilpua2011,winslow2015,lugaz2018,mishra2021,pal2023,palmerio2024}, interactions with other structures \citep[e.g.,][]{moestl2012,prise2015,winslow2016,winslow2021,lugaz2022}, physical effects such as erosion, deformation, distortion \citep[e.g.,][]{ruffenach2012,wang2018,palmerio2021,weiss2021} can induce fundamental changes in ICME measurements from one spacecraft to another and limit the generality of the findings. Another intrinsic drawback of most of the previous statistical approaches has been examining ICME evolution with clusters of in situ measurements at certain locations rather than a more continuous coverage. The addition of the Parker Solar Probe \citep[PSP;][]{fox2016,raouafi2023} and Solar Orbiter \citep[SolO;][]{muller2013,muller2020} missions to the Heliophysics System Observatory (HSO) improves upon the existing clusters of measurements at specific points in the inner heliosphere and provide more opportunities for multi-spacecraft encounters as well.

\justify
In this study, we take advantage of the extensive radial spread of PSP observations to investigate ICME evolution from different perspectives. Before the PSP mission, such investigations of the evolution of the magnetic field and plasma quantities inside an ICME have been limited to a radial extent beyond 0.29~au only. With PSP now in the fold and providing data coverage of the innermost heliosphere below 0.29~au, this enables us to advance our current understanding of ICME evolution in the inner heliosphere.

\justify
This paper is organized as follows. In Section~\ref{sec:MD}, we introduce the data and method used for the identification of ICMEs. In Section~\ref{sec:Results}, we present the compiled PSP ICME list, statistical results, and brief discussions. In Section~\ref{sec:SC}, we summarize the findings and make our conclusions.

\section{Data and Methodology} \label{sec:MD}

\justify
The PSP mission was launched on 2018 August 12. The mission consists of 24 highly elliptical orbits and 7 Venus gravity assist flybys (used to gradually decrease the perihelion every few orbits). The perihelion of orbit 22 will reach 9.86 solar radii on 2024 December 24 \citep[][]{raouafi2023}. During the mission, PSP will provide unprecedented measurements of the solar wind closer to the Sun. Such measurements will offer crucial insights into creating a more complete scenario of the CME/ICME-related phenomena, from the genesis at the Sun to the inner heliosphere.

\justify
To compile the PSP ICME list, we use Level 2 magnetic field data with a 1-min cadence from the Fluxgate Magnetometer (MAG) part of the FIELDS \citep{bale2016} instrument and 28-sec moment averages of bulk solar wind measurements from the Solar Probe Cup \citep[SPC;][]{case2020}, which is part of the Solar Wind Electrons Alphas and Protons \citep[SWEAP;][]{kasper2016} instrument suite onboard PSP, in Radial Tangential Normal (RTN) coordinates \citep{franz2002}. Because the PSP ICMEs investigated in this study are all beyond 0.2~au (see Table~\ref{tab:list} in Section~\ref{sec:Results}), SPC data can describe the solar wind well compared to the Solar Probe ANalyzer-Ions \citep[SPAN-i;][]{livi2022} and Solar Probe ANalyzer-Electrons \citep[SPAN-e;][]{whittlesey2020} detectors. This is because SPC has a better field of view (FOV) beyond encounter mode, whereas SPAN has a better FOV during encounter mode, meaning that SPC generally has better measurements outside 0.25~au. In addition, we use only the good-quality SPC data (quality flag=0) filtered using the data quality indicator.


\justify
In this manuscript, the term ``ICME'' refers to the whole structure, including the shock (if present), sheath (if present), and magnetic ejecta (ME). The identification of ICMEs in PSP observations is not always straightforward. Unlike near/at 1~au, it is not possible to adopt a quantitative approach since the position of PSP is not stationary. Due to this, we also have to be careful about features that are artifacts of PSP moving closer to the Sun or farther out, rather than a part of an ICME. Therefore, we adopt a more qualitative approach, reliant upon visual identification. We classify an ICME if the corresponding measurements match the majority of the following identification criteria (for the ME) given the available data: (i) significant enhancement of the magnetic field, (ii) rotation in the magnetic field vector, (iii) coherence of the magnetic field (smooth rotations and fewer fluctuations), (iv) low proton density, (v) low proton thermal speed, and (vi) proportionate duration to distinguish between small-scale structures that are not ICMEs \citep[see e.g.,][]{moldwin2000,janvier2014}. However, due to the unavailability of good-quality SPC data for several events, the identification is primarily based on criteria (i)--(iii) and (vi). 

\begin{figure*}[htbp]
\centering
\includegraphics[width=0.90\linewidth]{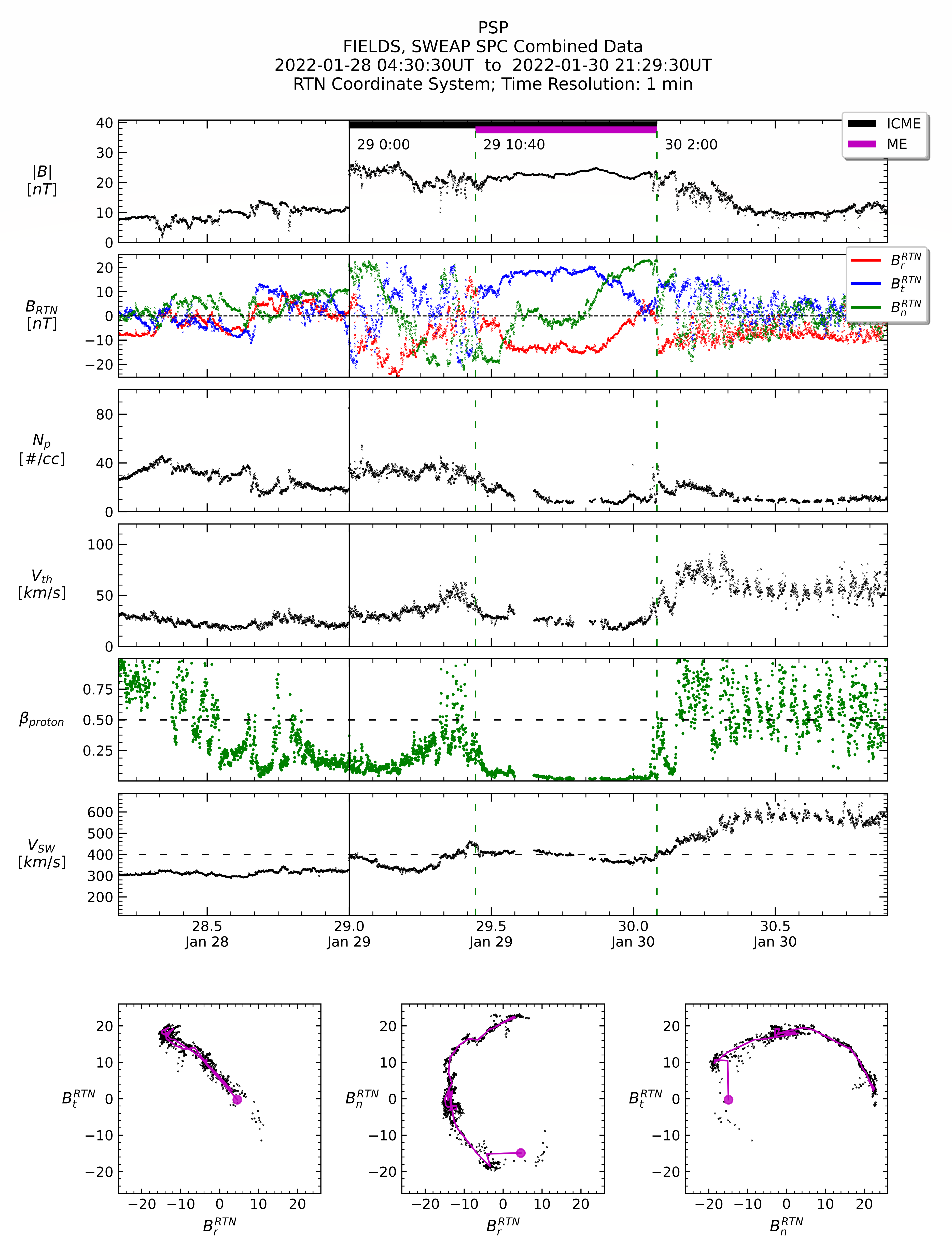}
\caption{PSP observations of an ICME on 2022 Jan 29 at 0.67~au. From top to bottom, the panels represent the total magnetic field strength, magnetic field vector in RTN coordinates, proton density, proton radial (most probable) thermal speed, proton beta, and proton bulk speed. The vertical solid black line indicates the shock arrival, the region between the solid black line and the first dashed green line is the sheath, and the region between the two dashed green lines is the ME. Magnetic hodograms, overlaid with smoothed, connected traces (magenta lines) in RTN coordinates for the ME duration (the magenta circle denotes the ME start) are shown in the three subplots at the bottom.}
\label{fig:PSP}
\end{figure*} 

\justify
Figure~\ref{fig:PSP} shows the magnetic field and plasma observations of an example ICME encountered by PSP. All the ICME identification criteria are satisfied by this event. The ME magnetic field strength is ${\sim}$100{\%} stronger compared to the upstream solar wind (defined as an 8-hr interval before the ICME arrival). The field rotations are smooth for the entire ME duration. The ME also exhibits low proton density and temperature. The duration of the ME is ${\sim}$15 hours. Although this ICME is not fast, at ${\sim}$400~km$\cdot$s$^{-1}$, it drives a shock and has an ${\sim}$10.7-hr sheath region.  

\section{PSP ICME List} \label{sec:Results}

\justify
Table~\ref{tab:list} lists the ICMEs that we identified from PSP observations based on the criteria mentioned in Section~\ref{sec:MD}. The ICME list has PSP observations ranging from 0.23~au to 0.83~au.

\begin{figure}[ht!]
\centering
\includegraphics[width=1\linewidth]{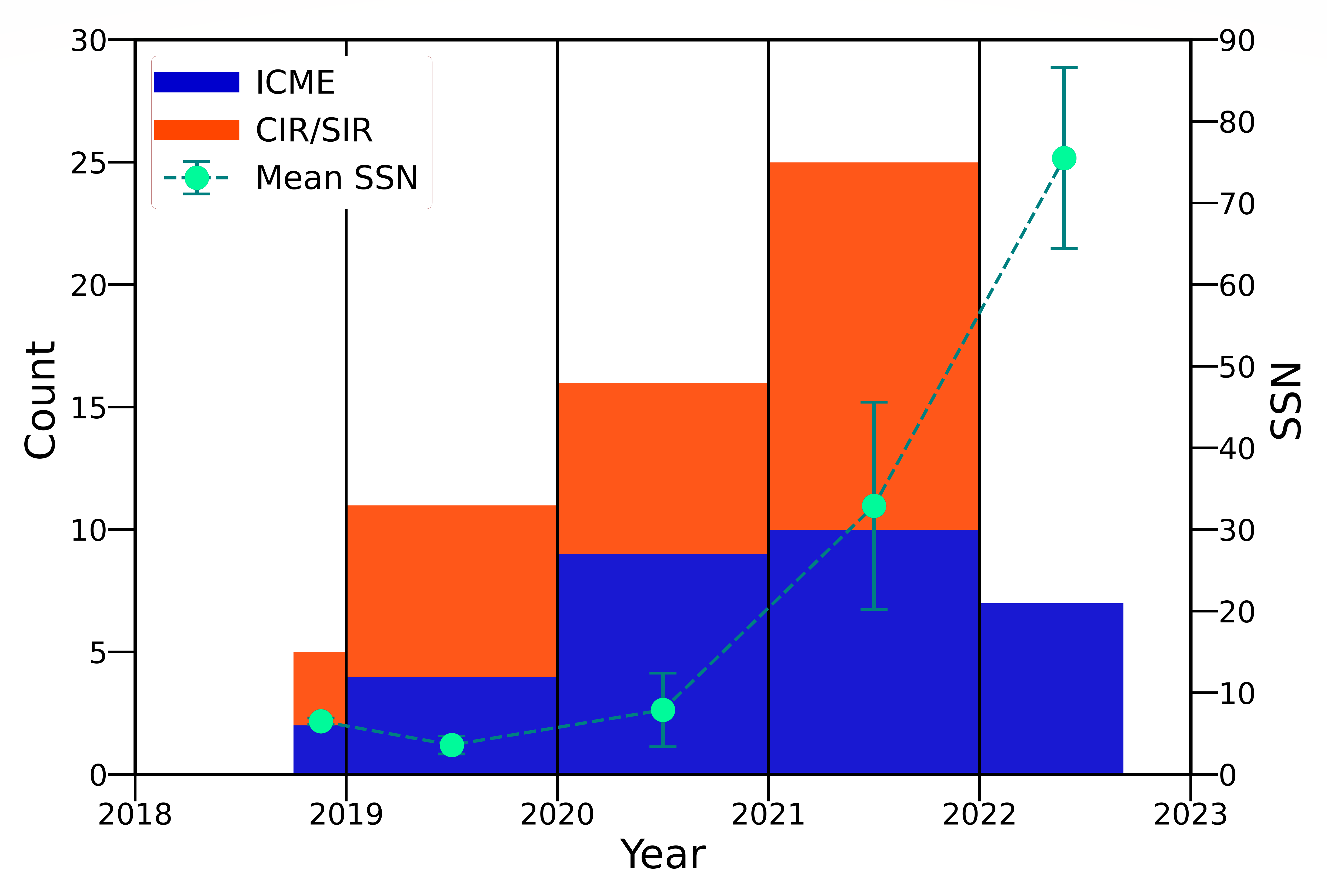}
\caption{Annual occurrences of ICMEs (in blue) and CIR/SIRs (in red) measured by PSP as a function of time. Green dots indicate the annual mean SSN with the error bars as the standard deviations. For the years 2018 and 2022, the ICME and CIR/SIR counts and mean SSN correspond to 3-month (from October to December) and 8-month (from January to August) periods respectively and are not normalized to whole years. The CIR/SIRs are listed from \citet{allen2021} and extend until the end of 2021.}
\label{fig:occurrence}
\end{figure}

\begin{deluxetable*}{cccccccc}[htbp]
\tablecaption{The PSP ICME list. Column~1 shows the chronological order of events. Column~2 represents PSP's heliocentric distance (\textit{r}) at the onset of the ICME arrival. Column~3 lists the shock (for events with both shock and sheath)/sheath (for events with no shock but a sheath)/ME (for events with no shock and sheath) arrival times. Column~4 and Column~5 highlight the ME start and end times respectively. Column~6 and Column~7 report the ME classes (see Subsection~\ref{ssec:complexity} for explanation) and average ME magnetic field strengths (B$_\textrm{ME}$) in order, whereas Column~8 has the distortion parameter (DiP, see Subsection~\ref{ssec:distortion} for details) values.
\label{tab:list}}
\tablehead{
\colhead{Event No} & \colhead{\textit{r}} & \colhead{ICME Start Time} & \colhead{ME Start Time} & \colhead{ME End Time} & \colhead{ME Class} & \colhead{B$_\textrm{ME}$} & \colhead{DiP} \\
\colhead{} & \colhead{au} & \colhead{YYYY MM/DD hh:mm} & \colhead{YYYY MM/DD hh:mm} & \colhead{YYYY MM/DD hh:mm} & \colhead{} & \colhead{nT} & \colhead{}  \\
}
\startdata
1     & 0.28  & 2018 10/30 20:55 & 2018 10/30 20:55 & 2018 10/31 08:10 & F- & 52.11 & 0.56 \\
2     & 0.23  & 2018 11/11 23:51 & 2018 11/11 23:51 & 2018 11/12 06:07 & F- & 82.08 & 0.53 \\
3     & 0.55  & 2019 03/15 09:00 & 2019 03/15 12:14 & 2019 03/15 17:45 & F- & 31.30 & 0.52 \\
4     & 0.41  & 2019 03/23 19:15 & 2019 03/23 19:15 & 2019 03/24 17:45 & F- & 25.98 & 0.15 \\
5     & 0.80   & 2019 10/13 19:03 & 2019 10/13 22:48 & 2019 10/14 21:07 & F$_\textrm{r}$ & 13.27 & 0.40 \\
6     & 0.83  & 2019 12/16 18:04 & 2019 12/16 18:04 & 2019 12/17 08:03 & E$_\textrm{j}$  & 13.37 & 0.49 \\
7     & 0.34  & 2020 01/20 19:00 & 2020 01/20 19:00 & 2020 01/21 06:30 & F- & 41.55 & 0.53 \\
8     & 0.41  & 2020 02/11 05:08 & 2020 02/11 05:08 & 2020 02/11 11:45 & F$_\textrm{r}$ & 40.05 & 0.52 \\
9     & 0.36  & 2020 05/28 09:20 & 2020 05/28 09:20 & 2020 05/28 14:50 & F$_\textrm{r}$ & 35.46 & 0.51 \\
10    & 0.45  & 2020 06/22 20:40 & 2020 06/23 03:20 & 2020 06/23 16:50 & E$_\textrm{j}$  & 15.74 & 0.55 \\
11    & 0.51  & 2020 06/25 11:53 & 2020 06/25 15:59 & 2020 06/26 08:15 & F- & 19.63 & 0.41 \\
12    & 0.48  & 2020 09/12 10:23 & 2020 09/12 13:33 & 2020 09/12 19:35 & E$_\textrm{j}$  & 15.47 & 0.44 \\
\textcolor{gray}{13}  & \textcolor{gray}{0.69}  & \textcolor{gray}{2020 10/27 04:50} & \textcolor{gray}{2020 10/27 07:15} & \textcolor{gray}{2020 10/27 15:40} & \textcolor{gray}{-} & \textcolor{gray}{-}     & \textcolor{gray}{-} \\
14    & 0.81  & 2020 11/29 23:07 & 2020 11/30 03:21 & 2020 11/30 16:26 & C$_\textrm{x}$ & 10.75 & 0.50 \\
15    & 0.81  & 2020 11/30 18:35 & 2020 12/01 02:24 & 2020 12/01 11:17 & F$_\textrm{r}$ & 28.39 & 0.40 \\
16    & 0.62  & 2021 02/11 16:18 & 2021 02/11 16:18 & 2021 02/12 03:56 & C$_\textrm{x}$ & 12.16 & 0.42 \\
\textcolor{gray}{17}  & \textcolor{gray}{0.64}  & \textcolor{gray}{2021 02/12 11:18} & \textcolor{gray}{2021 02/12 11:18} & \textcolor{gray}{-}     & \textcolor{gray}{-} & \textcolor{gray}{-}     & \textcolor{gray}{-} \\
18    & 0.68  & 2021 05/28 03:00 & 2021 05/28 08:10 & 2021 05/29 12:04 & F$_\textrm{r}$ & 18.36 & 0.18 \\
\textcolor{gray}{19}  & \textcolor{gray}{0.69}  & \textcolor{gray}{2021 05/30 08:08} & \textcolor{gray}{2021 05/30 14:18} & \textcolor{gray}{2021 06/01 02:50} & \textcolor{gray}{-} & \textcolor{gray}{-}     & \textcolor{gray}{-} \\
20    & 0.76  & 2021 06/10 04:38 & 2021 06/10 16:52 & 2021 06/11 08:54 & C$_\textrm{x}$ & 11.34 & 0.45 \\
21    & 0.77  & 2021 06/12 03:40 & 2021 06/12 07:20 & 2021 06/13 01:12 & F+ & 19.98 & 0.53 \\
22    & 0.76  & 2021 06/30 14:58 & 2021 06/30 22:10 & 2021 07/01 17:12 & F+ & 25.75 & 0.31 \\
\textcolor{gray}{23}  & \textcolor{gray}{0.69}  & \textcolor{gray}{2021 07/11 10:25} & \textcolor{gray}{2021 07/11 10:25} & \textcolor{gray}{2021 07/11 23:25} & \textcolor{gray}{-} & \textcolor{gray}{-}     & \textcolor{gray}{-} \\
24    & 0.78  & 2021 09/26 08:56 & 2021 09/26 23:04 & 2021 09/27 12:50 & F- & 14.39 & 0.48 \\
25    & 0.44  & 2021 11/09 16:22 & 2021 11/09 18:40 & 2021 11/10 04:20 & F$_\textrm{r}$ & 53.89 & 0.50 \\
26    & 0.67  & 2022 01/29 00:00 & 2022 01/29 10:40 & 2022 01/30 02:00 & F$_\textrm{r}$ & 22.42 & 0.51 \\
27    & 0.58  & 2022 02/05 08:40 & 2022 02/05 08:40 & 2022 02/05 14:00 & F- & 14.51 & 0.56 \\
28    & 0.38  & 2022 02/16 07:25 & 2022 02/16 15:18 & 2022 02/17 06:32 & F- & 45.88 & 0.43 \\
29    & 0.52  & 2022 03/14 16:30 & 2022 03/14 16:30 & 2022 03/16 16:00 & E$_\textrm{j}$  & 15.46 & 0.46 \\
30    & 0.69  & 2022 07/02 04:40 & 2022 07/02 18:30 & 2022 07/04 16:30 & E$_\textrm{j}$  & 15.50 & 0.47 \\
31    & 0.76  & 2022 07/18 22:30 & 2022 07/18 22:30 & 2022 07/19 18:00 & C$_\textrm{x}$ & 18.29 & 0.47 \\
32    & 0.56  & 2022 08/18 13:30 & 2022 08/18 16:00 & 2022 08/19 00:00 & F$_\textrm{r}$ & 56.46 & 0.44 \\
\enddata
\end{deluxetable*}

\justify
The ICME list covers the period from 2018 October to 2022 August. This period includes the minimum and rising phases of the current solar cycle (SC), SC25. Figure~\ref{fig:occurrence} shows ICMEs and co-rotating/stream interaction regions (CIR/SIRs) measured by PSP. The CIR/SIR counts are from the compiled list of \citet{allen2021}. The plot shows a clear rise in the annual ICME occurrences from the solar minimum to the rising phase. The number of ICMEs has risen from 4 events in 2019 to 10 in 2021. The CIR/SIR occurrences also follow a similar progression. In Figure~\ref{fig:occurrence}, we also plot the annual mean sunspot number (SSN) from the Solar Influences Data Analysis Center (SIDC\footnote{\url{http://sidc.be/silso/home}}) for the years 2018--2022. The annual mean is computed from the 13-month smoothed monthly total SSN. Previous studies have found good correlations between the ICME rate and SSN \citep[e.g.,][]{jian2006,gopalswamy2010,jian2018,li2018,moestl2020}. Similarly, for this period as well, the SSN exhibits positive relationships with the annual ICME and CIR/SIR occurrences. 



\justify
Due to the unique orbits of PSP, we cannot do the usual averages as we also have to incorporate the radial spread in observations. Thus, we separate the ICMEs into two radial blocks. The first block consists of 14 ICMEs from 0.23~au to 0.56~au, whereas the second block also consists of 14 ICMEs but from 0.58~au to 0.83~au. The average ME magnetic field strengths (B$_\textrm{ME}$) in the first and second blocks are 37.9$\pm$19.3~nT and 17.0$\pm$5.4~nT respectively. The lower average field strength (by ${\sim}$55{\%}) in the second block is consistent with the decrease in average values reported in previous statistical studies \citep[e.g.,][]{winslow2015,janvier2019,salman2020a}. We also perform Welch's t-test to measure the statistical difference between the average values. We find the difference between the average field strengths of the two populations to be statistically significant with 95{\%} confidence (\textit{p}-value=\textbf{0.001}).   

\justify
In total, we list 32 ICMEs measured by PSP. However, our statistical analyses presented in the following subsections are based on 28 ICMEs (excluding events 13, 17, 19, and 23 which are marked by gray fonts in Table~\ref{tab:list}) rather than 32 since these four ICMEs have considerable MAG data gaps within respective ME durations. 


\subsection{ME Complexity} \label{ssec:complexity}

\justify
The common consensus in the heliophysics community is that CMEs erupt with a flux-rope (FR) configuration \citep[e.g.,][]{chen1996,bothmer1998,green2018}, where nested, helical magnetic field lines are wrapped around a central axis \citep[][]{lundquist1951,goldstein1983}. As the transition from CME to ICME occurs and during ICME propagation, different physical phenomena can alter this FR configuration. In addition, such changes in coherent configurations can also be attributed to spacecraft crossings farther from FR centers \citep[e.g.,][]{jian2006,kilpua2011} and interactions with other large-scale structures (i.e., high-speed streams, CIR/SIRs, the heliospheric current/plasma sheet).

\justify
To investigate the radial evolution of ME configurations, we use the morphological classification scheme of \citet{nieveschinchilla2018,nieveschinchilla2019}. In this scheme, magnetic hodograms are divided into five categories based on rotation in the magnetic field vector within the ME duration. The categories are: (i) F- is a single rotation less than 90$^{\circ}$, (ii) F$_\textrm{r}$ is a single rotation between 90$^{\circ}$--180$^{\circ}$, (iii) F+ is a single rotation greater than 180$^{\circ}$, (iv) C$_\textrm{x}$ includes multiple rotations, and (v) E$_\textrm{j}$ represents an unclear rotation. The categories represent different ME topologies and/or spacecraft trajectories. F- and F$_\textrm{r}$ populations can arise from large and small impact parameter crossings respectively \citep[][]{nieveschinchilla2019}. F+ configurations may correspond to MEs with significant curvatures or complex topologies like spheromaks \citep[][]{vandas1997,nieveschinchilla2019,scolini2021} or double FRs \citep[][]{lugaz2013,osherovich2013,nieveschinchilla2019}. C$_\textrm{x}$ class can be associated with eruptions from complex active regions at the Sun \citep[][]{nieveschinchilla2019} or interactions between multiple ICMEs \citep[][]{lugaz2017}. E$_\textrm{j}$ group may be indicative of spacecraft crossings through ME legs with untwisted magnetic field lines \citep[][]{owens2016,scolini2022}.

\begin{figure}[htbp]
\centering
\includegraphics[width=1\linewidth]{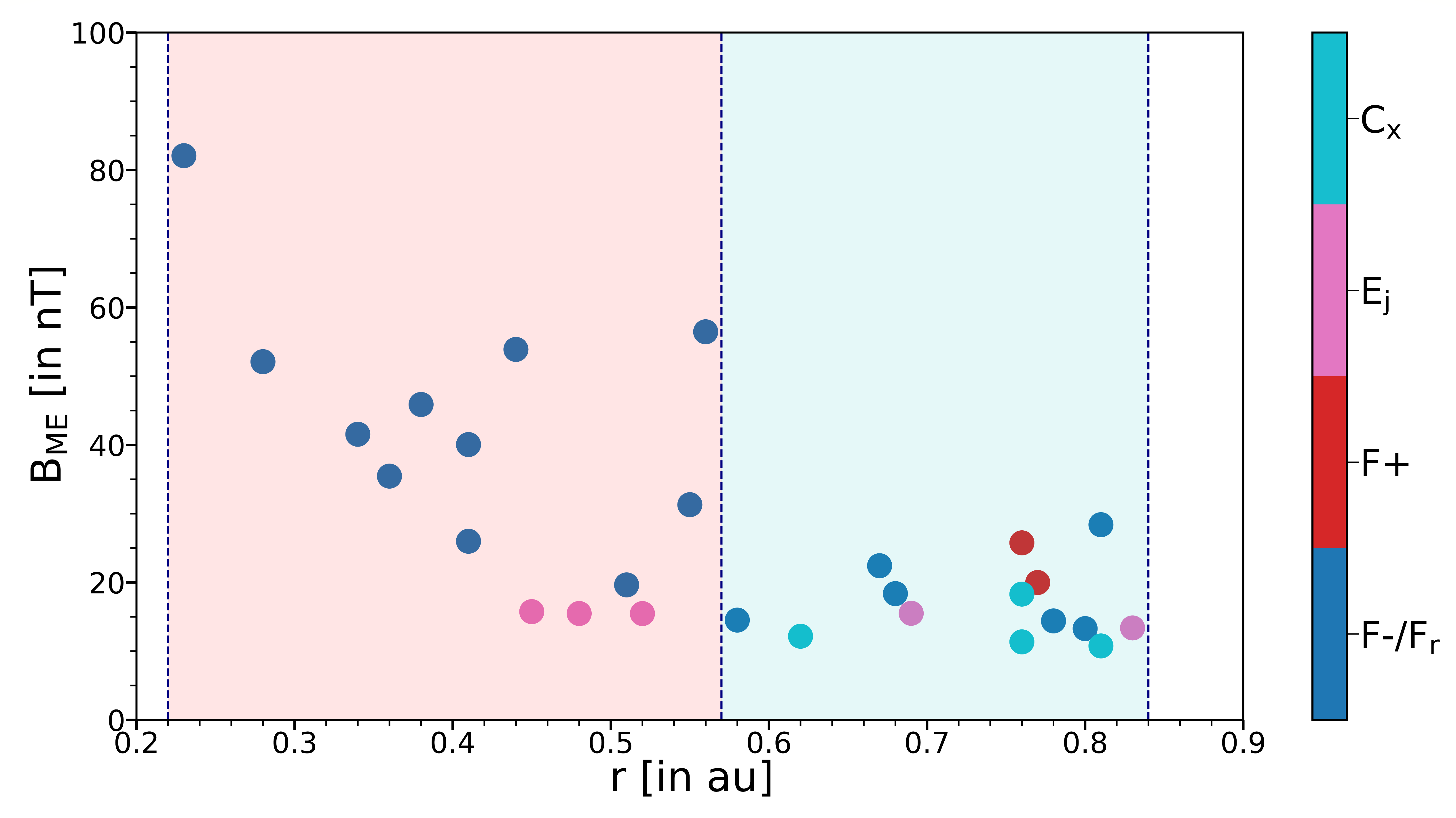}
\caption{Distribution of average ME magnetic field strength (in nT) as a function of heliocentric distance (in au). The ICMEs are color-coded based on the classification scheme of \citet{nieveschinchilla2018,nieveschinchilla2019}. The two shaded regions correspond to the two radial blocks (see text for details).}
\label{fig:complexity}
\end{figure}

\justify
We now examine how the ME configurations evolve with heliocentric distance (\textit{r}, see Figure~\ref{fig:complexity}). From our observations of 28 ME configurations, we find that 17 (61{\%}) are F- or F$_\textrm{r}$, 2 (7{\%}) are F+, 4 (14{\%}) are C$_\textrm{x}$, and 5 (18{\%}) are E$_\textrm{j}$. Also, we see that closer to the Sun, most of the ME configurations are either F- or F$_\textrm{r}$, whereas more complex configurations like C$_\textrm{x}$ and E$_\textrm{j}$ are observed at larger \textit{r}. Another important finding is that in a similar radial range, weaker ICMEs tend to be more complex.

\justify
However, it is also important to mention how PSP's longitudinal trajectory and relative speed can lead to observations of different magnetic signatures within a single encounter. Closer to the Sun, the significantly faster speed of PSP could mean that a CME is crossed twice. At 0.23~au, for specific instances, PSP's speed of ${\sim}$72~km$\cdot$s$^{-1}$ can be more than 10{\%} of a CME speed and may affect the observations of field rotations. As an example, \citet{moestl2020} modeled such an encounter at \textless0.1~au to demonstrate how the crossings of two different portions of the CME (apex and leg) could result in differences in field rotations. \citet{moestl2020} also highlighted the importance of CME parameters (i.e., speed, shape) and FR inclination on the double-crossing as a flatter cross-section would increase the probability of such an encounter. 


\justify
This finding is another illustration that ME complexity can increase with \textit{r} \citep[e.g.,][]{riley2013}. The same was identified by \citet{scolini2022} as well from a statistical analysis of 31 ICMEs listed in \citet{salman2020a} that are observed in radial alignment between 0.3 and 1~au. \citet{scolini2022} found that interactions with other large-scale structures are the primary drivers of increased ME complexities. \citet{richardson2010} in a previous study indicated that the occurrence of complex configurations is also strongly dependent on the SC phase since the probability of an ICME interacting with other large-scale structures and losing its coherent configuration is greater during solar maxima than minima.

\justify
We also examine how the solar wind is impacted following an ICME transit. Even after the ICME passage, the post-ICME solar wind goes through a relaxation period to return to the same level of magnetic field strength as the pre-ICME solar wind \citep[see][]{janvier2019}. At 1~au, this period can extend up to 2 to 5 days, as pointed out by \citet{temmer2017}. \citet{carcaboso2020} observed a notable amount of bidirectional suprathermal electrons (BDE) in the post-ICME solar wind at 1~au where the post-ICME region was defined as 1.2 times the ICME duration. Herein, we define an interval equivalent to the ME duration as the post-ICME region. Then, we determine the percentage increase/decrease in the post-ICME magnetic field strength compared to its pre-ICME value. We define an 8-hr interval before the ICME arrival as the pre-ICME region. Due to the nature of PSP's transitory orbit, we assign a scaling factor $r^{2}$ to the pre-ICME and post-ICME values. This normalizes the effect of \textit{r} on the average values. We find that in the first radial block (0.23--0.56~au), the post-ICME magnetic field strength is 13{\%} stronger on average compared to its pre-ICME value. However, in the second radial block (0.58--0.83~au), we see that the post-ICME magnetic field strength is on average ${\sim}$44{\%} stronger compared to the pre-ICME value.

\justify
Again, similar to how PSP's trajectory and speed can affect how we see field rotations, this finding of the post-ICME magnetic field being significantly more impacted for the second radial block can be a manifestation of the nature of PSP's orbit. This is because, in the first or closest radial block, we have PSP observations when the spacecraft is moving inwards more often, whereas, for the second or farthest radial block, we have a nearly equal number of inbound and outbound observations.

\subsection{ICME Magnetic Field}

\justify
Expansion is an integral component of ICME evolution as it governs parameters such as the ME magnetic field strength, density, and size. Previous studies have found that the expansion rate is greater within 1~au than beyond 1~au \citep[e.g.,][]{leitner2007,davies2021b,davies2022}. Even within the 1~au range, the expansion is known to be occurring at a higher rate closer to the Sun due to the strong initial ME magnetic pressure, sometimes causing overexpansion \citep[e.g.,][]{gosling1994,gosling1998} and then this rate decreases gradually as the ICME travels through interplanetary space \citep[e.g.,][]{farrugia2008,vrsnak2019,lugaz2020a}. However, the ICME expansion rate in interplanetary space can also significantly vary within a wide range, depending on the pressure balance with the background solar wind and possible interactions \citep[e.g.,][]{luhmann2020,gopalswamy2022}.

\begin{figure}[htbp]
\centering
\includegraphics[width=1\linewidth]{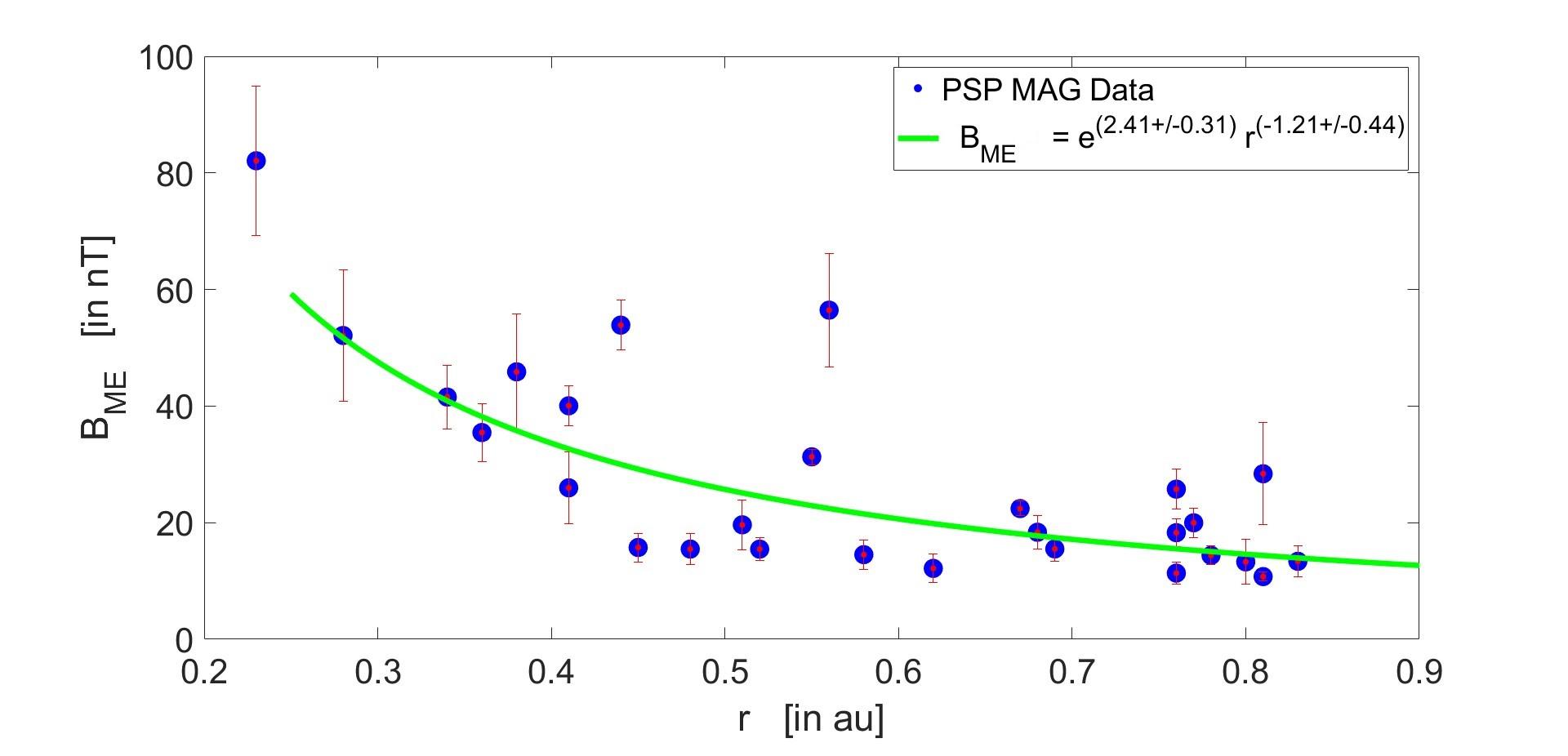}
\caption{Average ME magnetic field strength (in nT) from PSP observations plotted as a function of heliocentric distance (in au). The power-law curve (in green) represents the best-fit curve to our data set. The error bars (in red) represent the standard deviations.}
\label{fig:radialB}
\end{figure}

\justify
ME properties such as magnetic field strength can be used as a deterministic feature to determine its radial evolution. The variation of the ME magnetic field strength with \textit{r} is also considered an indirect measure of its global expansion. Based on theoretical assumptions and statistical fits, the magnetic field strength is expected to fall off as a power-law with \textit{r} \citep[e.g.,][]{farrugia2005,wang2005,leitner2007,demoulin2009,gulisano2010,winslow2015,good2019,vrsnak2019,salman2020a,davies2021b,davies2022}. To examine this, we adopt a similar technique as \citet{winslow2015,salman2020a}. We perform a multi-linear robust regression in logarithmic space to fit a power law curve to our data set (see Figure~\ref{fig:radialB}). The basis of this fitting technique is an iterative re-weighted least squares approach with a bi-square weighting function to ensure that less weight is attributed to outliers than in ordinary least squares fitting. From this, we find the relationship, B$_\textrm{ME}$= $11.13^{+4.05}_{-2.96}$ $r^{(-1.21\pm0.44)}$. The uncertainties assigned to the power-law index represent a 95{\%} confidence interval.

\justify
This relationship between B$_\textrm{ME}$ and \textit{r} indicates a lower ME expansion rate than previous studies. Using Helios 1 and 2 measurements, \citet{gulisano2010} found this power-law index to be -1.85$\pm$0.07, whereas from MErcury Surface, Space ENvironment, GEochemistry, and Ranging \citep[MESSENGER;][]{solomon2001} and Advanced Composition Explorer \citep[ACE;][]{stone1998} measurements, \citet{winslow2015} reported this power-law index to be -1.95$\pm$0.19. In a recent study, \citet{davies2022} determined this power-law index to be -1.49$\pm$1.12 within the 1~au range. 

\justify
However, it is also important to note that statistical fits can mask the behavior of individual events \citep[see discussion in][]{farrugia2005,good2019,vrsnak2019,salman2020a,davies2022}. The large spread in radial dependencies, seen from the limits of the power-law index (-1.65 to -0.77) is a possible reflection of such ICME-to-ICME variability. A similarly large spread, $r^{(-1.34\pm0.71)}$ was also reported in \citet{good2019} when the power-law index was calculated for individual events. In a similar approach, taking advantage of multi-spacecraft measurements, \citet{salman2020a} calculated the exponential decrease of the peak ME magnetic field strength with \textit{r} for individual ICMEs and made a comparison with the derived statistical relationship. \cite{salman2020a} found that 37 out of the 45 ICMEs showed a rate of decrease outside of the 95{\%} confidence interval of the statistical fit. In a recent multi-spacecraft study, \citet{davies2022} also calculated the average power-law index for individual events observed within 1~au and found the limits of the index to have noticeable differences compared to the limits of the index derived from a statistical fit.

\justify
We also investigate the magnetic field fluctuations within ICME sheaths \citep[e.g.,][]{masias2016,kilpua2019,kilpua2020,regnault2020,salman2020b,kilpua2021,salman2021}. Such fluctuations are a consequential feature of different physical mechanisms (i.e., shock compression, field line draping) in action \citep[e.g.,][]{mcComas1988,neugebauer1993,kataoka2005}. Previous studies have also investigated fluctuations in different portions of the sheath \citep[see][]{kilpua2019,kilpua2020}. Here, we consider the fluctuations as a whole. To quantify magnetic field fluctuations within sheaths, we adopt a similar approach as \citet{salman2020b,salman2021}. We introduce a total root-mean-square [B$_\textrm{rms(tot)}$] that is the sum of root-mean-square-deviations for each 5-min time interval (B$_\textrm{rms}$), as defined in Equation \ref{eq:1} and normalized by the average sheath magnetic field strength. 

\begin{equation}\label{eq:1}
\text{B$_\textrm{rms}$}=\sqrt{\sum_{i=1}^{n} \frac{(B_{i} - \langle B \rangle)^2}{n}\\}
\end{equation}

\justify
Here, B$_{i}$ (i=1...5) and $\langle$B$\rangle$ represent the magnetic field strength for each 1-min time step and the temporal average for 5-min intervals respectively, and n is the number of measurements.

\justify
We now explore correlations between B$_\textrm{rms(tot)}$ with $r$, sheath duration (T$_\textrm{sheath}$), and the ratio of the average ME magnetic field strength to the upstream magnetic field (B$_\textrm{ME}$/B$_\textrm{SW}$, here B$_\textrm{SW}$ is the average magnetic field strength of an 8-hr interval before the ICME arrival). We find no correlation between B$_\textrm{rms(tot)}$ and $r$ (Pearson's correlation coefficient or PCC=0.12 with \textit{p}-value=0.657) and B$_\textrm{rms(tot)}$ and T$_\textrm{sheath}$ (PCC=-0.14 with \textit{p}-value=0.616). However, we report a strong positive correlation between B$_\textrm{rms(tot)}$ and B$_\textrm{ME}$/B$_\textrm{SW}$ (PCC=\textbf{0.76} with \textit{p}-value=\textbf{0.001}).

\justify
The nonexistent correlation between B$_\textrm{rms(tot)}$ and $r$ is surprising. The evolution of the sheath involves continuous accumulation of upstream solar wind plasma and magnetic field. This accumulation occurs at different \textit{r} locations during ICME propagation. Therefore, the layers of the sheath correspond to a heterogeneous plasma and magnetic field distribution \citep[e.g.,][]{kaymaz2006,siscoe2007}, with various possible confined waves, discontinuities, and reconnection exhausts \citep[e.g.,][]{kataoka2005,ala-lahti2018}. For this reason, a positive correlation between B$_\textrm{rms(tot)}$ and $r$ can be expected since with increasing $r$, the time-scale of the accumulation process will increase, which can result in increased inhomogeneity within the sheath, leading to possible increased fluctuations. A possible reason for this observed no correlation could be that there are processes in the solar wind itself (such as wave-particle interactions) to decrease the inhomogeneity.



\justify
On the other hand, we see a strong positive correlation between B$_\textrm{rms(tot)}$ and B$_\textrm{ME}$/B$_\textrm{SW}$. Using logistic regression, \citet{salman2021} found that B$_\textrm{ME}$ and M$_\textrm{pseudo}$ (defined as the ratio between the ME leading-edge speed in the solar wind frame to the characteristic speed of the solar wind) are influential factors behind the observed distinct speed profiles (i.e., constant, increasing, decreasing) within sheaths near 1~au. Previous studies have also shown strong ME speed-magnetic field correlations for magnetic clouds \citep[MCs;][]{burlaga1981,burlaga1982,klein1982} at 1~au \citep[e.g.,][]{owens2005} and for ICMEs that do not change ME configurations from one measuring spacecraft in the inner heliosphere to another \citep[][]{scolini2022}. Thus, it is possible that the quantity B$_\textrm{ME}$/B$_\textrm{SW}$ is coupled to the ME kinematics (propagation and expansion). The ME kinematics significantly contributes to the arrangement of the plasma layers within the sheath. The layers are not only stacks of simple compressed plasma in the radial direction as the layers can slide laterally as well by different amounts \citep[see][]{siscoe2008}. Therefore, the ordering of the compressed plasma layers within the sheath can be a measure of sheath fluctuations and related to the quantity B$_\textrm{ME}$/B$_\textrm{SW}$.

\subsection{Formation and Expansion of Sheath}

\justify
The formation and development of the sheath is a topic that is still not well understood. Remote-sensing observations and in situ composition measurements have provided great insights into possible distances for sheath formation \citep[e.g.,][]{deForest2013,kilpua2017,lugaz2020b,temmer2021}. The basis of such considerations is that as CMEs primarily consist of coronal material (cooler, denser material likely of chromospheric or photospheric origin can also be present occasionally), composition signatures, consisting of both coronal and compressed (also shocked if there is a preceding shock) solar wind materials in the sheath or traces of only solar wind material can be an indicator of the approximate sheath formation distance. 

\begin{figure}[htbp]
\centering
\includegraphics[width=1\linewidth]{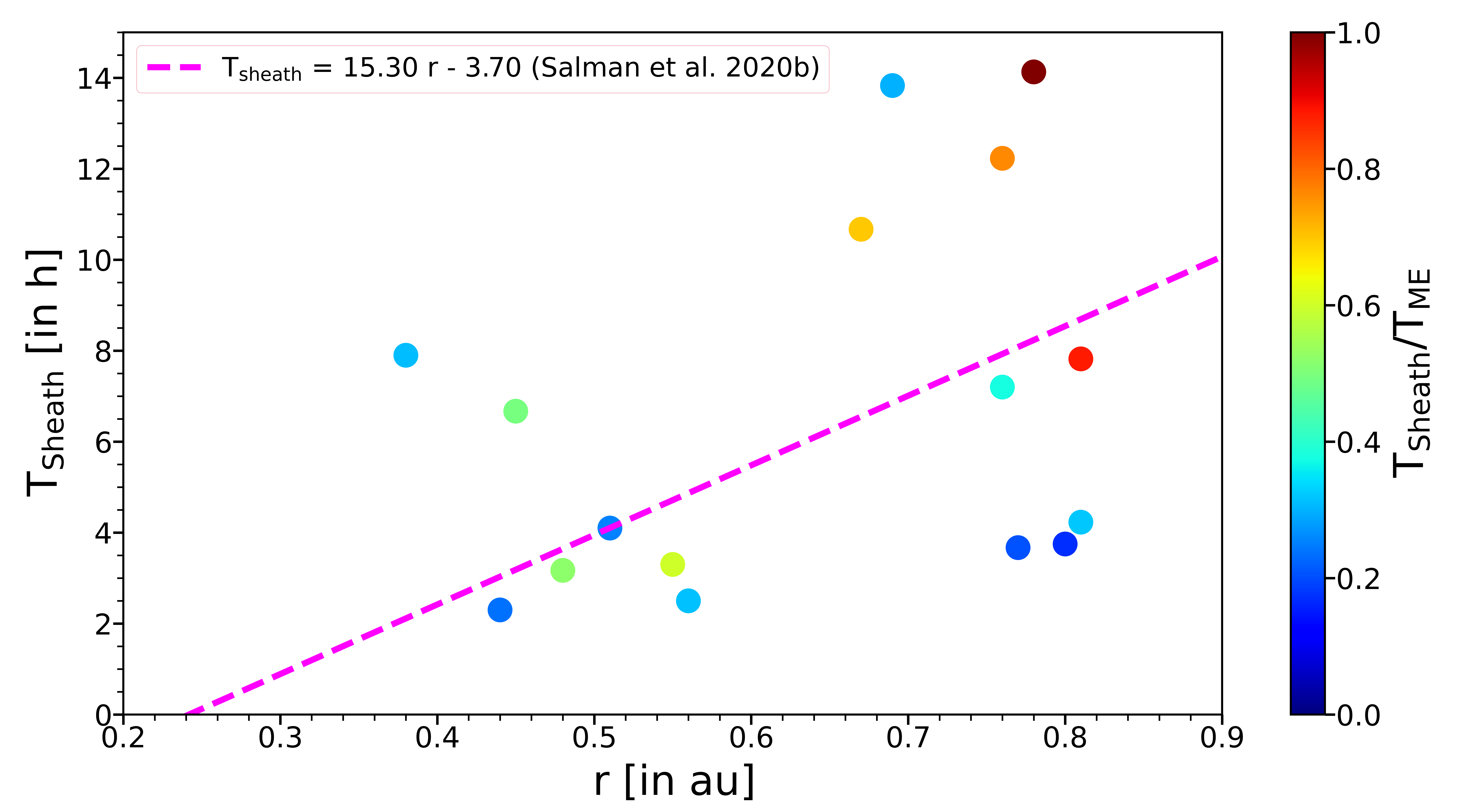}
\caption{Sheath duration (in h) plotted as a function of heliocentric distance (in au) with the ratio of sheath duration to ME duration as a second scale in the color bar. The magenta dashed line represents the best linear least squares fit reported in \citet{salman2020a}.}
\label{fig:sheathduration}
\end{figure}

\justify
In our PSP ICME list, there are 16 ICMEs with a well-defined sheath. From a linear approximation, \citet{salman2020a} found the sheath formation to start around 0.24~au. In a recent study, from a statistically derived density evolution over \textit{r} using Helios and PSP observations, \citet{temmer2022} theorized the sheath to start forming even before that, at ${\sim}$0.06~au. However, we observe that only 2 out of the 8 ICMEs in our list within 0.44~au have sheaths and none before 0.38~au. This finding can be a manifestation of our sampling technique as well. Both magnetic field and plasma measurements are required for the identification of sheaths with precision. Since for the majority of the mentioned 16 ICMEs, the sheaths are identified based on magnetic field measurements only (due to the lack of good-quality plasma data), this could have led to the non-identification of some sheaths. However, this finding can also hint at how the ME magnetic field strength and expansion closer to the Sun play an important role in sheath formation.

\justify
The sheath formation has contributions from both the ME propagation, where the solar wind is deflected at the nose, and expansion, which prevents the deflected solar wind from going around the ICME body \citep[see][]{siscoe2008}. Closer to the Sun, the expansion occurs because of the strong internal ME magnetic pressure. Since the period of this study primarily covers the minimum and early rising phase of SC25, the ICMEs are expected to be weaker in terms of magnetic field strengths. This means that expansion closer to the Sun may be occurring at a lower rate than usual. This will make it easier for the deflected solar wind at the nose of the ICME to go around, ceasing the possible pile-up of solar wind plasma in front of the ME.

\justify
In SC24, the ME expansion was generally stronger than expected for weaker ICMEs due to an evident drop in the magnetic field and heliospheric pressure \citep[e.g.,][]{gopalswamy2015,jian2018}. So, this hypothesis of weaker ICME expansion needs to be further evaluated by including ICMEs from the maximum phase of SC25. In addition, local ICME expansion from plasma measurements will provide knowledge regarding whether this expansion behavior persists throughout the inner heliosphere, as it did in SC24.

\justify
After this, we investigate the expansion of the sheath. Due to expansion, the quantity T$_\textrm{sheath}$ is expected to increase with \textit{r} during ICME propagation \citep[e.g.,][]{janvier2019,salman2020a}. \citet{salman2020a} performed a linear least squares fitting of sheaths observed in radial alignment between 0.3 and 1~au for the periods 2006--2013 and 2011--2015 and reported a correlation coefficient of \textbf{0.78} between T$_\textrm{sheath}$ and \textit{r} for a larger set of events. However, the correlation seen between T$_\textrm{sheath}$ and \textit{r} from our observation (see Figure~\ref{fig:sheathduration}) is poor (PCC=0.35 with \textit{p}-value=0.185). We also examine how the ratio of the duration of the sheath over that of the ME evolves with \textit{r}. This ratio was found to increase from MESSENGER to ACE by \citet{janvier2019}. But we also find a poor correlation between this pair of parameters (PCC=0.28 with \textit{p}-value=0.293). A possible reason for the identified poor correlations in our case compared to previous studies is the small number of events and the observed event-to-event variability. In our data set, we have a few sheaths that are already 6--8 hours long within 0.45~au, comparable to the average sheath duration of 8--9 hours near 1~au \citep[see][]{salman2020b}. On the other hand, we have a few sheaths that are only 3--4 hours long around 0.80~au. Such discrepancies are also possible manifestations of spacecraft crossings as the sheath duration has been shown to increase from the nose of the ICME toward the flanks in previous studies \citep[e.g.,][]{kilpua2017,salman2020a}.   




\subsection{Distortion of ME Magnetic Field} \label{ssec:distortion}

\justify
The ME expansion, combined with the global motion can lead to distortion of the observed ME magnetic field \citep[e.g.,][]{farrugia1995,demoulin2008,masias2016,janvier2019,demoulin2020,regnault2023}. From superposed epoch analysis at 1~au, \citet{masias2016} found the magnetic field strength profiles inside MCs to be strongly asymmetric, with the peak shifted to the front of the MC. This can be attributed to the time-delay factor (termed the ``aging'' effect) as the spacecraft encounters the front of the ICME structure earlier than the rear. As a result, since expansion affects the time evolution of the measured magnetic field, the weakening of the magnetic field during the spacecraft encounter will result in a decrease in magnetic field strength from the front to the rear. This will correspond to an apparent asymmetry and observed compression of the magnetic field near the front of the structure.  

\justify
Previous studies have attempted to quantify this asymmetry of the magnetic field within an ME \citep[e.g.,][]{nieveschinchilla2018,janvier2019,demoulin2020,lanabere2020}. In this study, we follow the mathematical formulation of \citet{nieveschinchilla2018} and determine the quantity DiP from the temporal average of the ME magnetic field strength. The DiP value is the fraction of the total ME duration where 50{\%} of the total magnetic field is accumulated. From this definition, DiP values range between 0 to 1 and provide a quantitative measure of the nature and extent of the asymmetry. DiP values \textless0.5 and \textgreater0.5 correspond to compression at the front and rear respectively. A DiP value of 0.50$\pm$0.07 corresponds to a symmetric magnetic field profile \citep[][]{nieveschinchilla2018}, meaning no significant amount of front or rear compression. The DiP value will approach 0 for higher compression at the front and 1 for higher compression at the rear.

\begin{figure}[t!]
\centering
\includegraphics[width=1\linewidth]{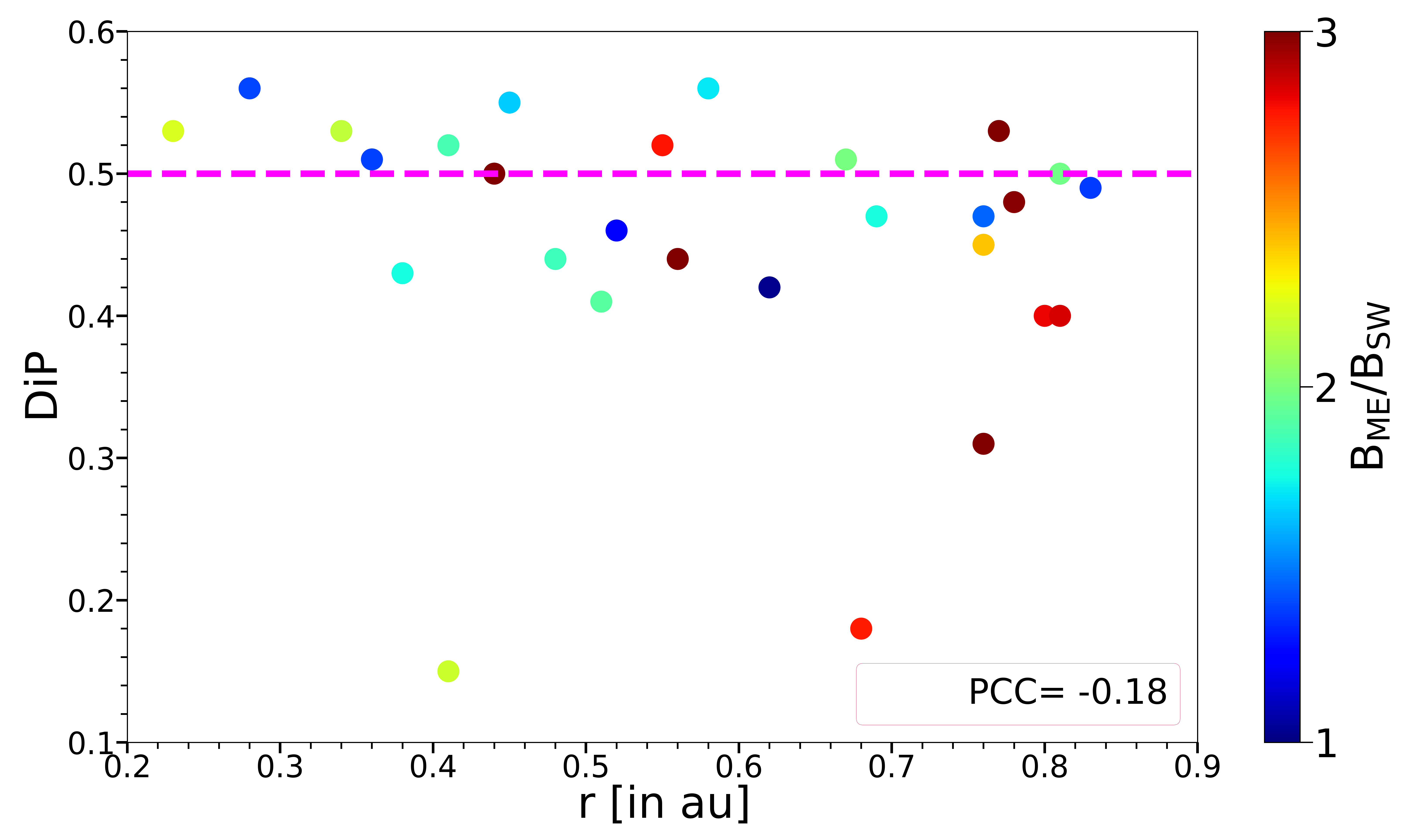}
\caption{DiP values plotted as a function of heliocentric
distance (in au) with the ratio of the average ME magnetic field strength to the upstream magnetic field as a second scale in the color bar. The dashed magenta line represents a DiP value of 0.5 that corresponds to a symmetric ME magnetic field profile.}
\label{fig:dip}
\end{figure}

\justify
Figure~\ref{fig:dip} shows the radial distribution of DiP values. Examining the DiP values, we find the range between 0.15--0.56. 18 of the MEs have DiP values \textless0.5, whereas 10 have DiP values \textgreater0.5. However, a closer look at the values reveals that 50{\%} of the DiP values lie within the range between 0.45--0.55, indicating approximately symmetric magnetic field profiles. This is in agreement with \citet{janvier2019} who found that slower ME profiles (like most of the ICMEs in our list) are more symmetric as the MEs relax with increasing $r$. 

\justify
We also find no correlation between DiP values and $r$ (PCC=-0.18 with \textit{p}-value=0.344). \citet{janvier2019} found that at MESSENGER, magnetic field profiles within MEs are more peaked toward the front. However, we do not observe a similar trend within Mercury’s orbit (0.31--0.47~au). \citet{lugaz2020a} theorized that the front of the ME can expand relatively freely in the presence of a sheath. In such a scenario, the front portion of the ME exhibits more expansion compared to the rear, resulting in the peak being shifted more toward the front. Since in our list, only 3 out of the 9 ICMEs within 0.47~au have sheaths, the constrained nature of ME front expansion can be a possible reason for the observed distribution of DiP values within Mercury’s orbit where most of the DiP values are around or above the 0.50 line (see Figure~\ref{fig:dip}). 

\justify
We also investigate two other correlations: whether the level of enhancement in the ME magnetic field (B$_\textrm{ME}$/B$_\textrm{SW}$) and ME duration (T$_\textrm{ME}$) have influence on the nature of asymmetries. We find no correlation between DiP values and B$_\textrm{ME}$/B$_\textrm{SW}$ (PCC=-0.25 with \textit{p}-value=0.208). For the second inspection, we only consider DiP values \textless0.5. However, we also find no correlation between DiP values and T$_\textrm{ME}$ (PCC=-0.14 with \textit{p}-value=0.587). ME expansion has consequences on the measured magnetic fields and a correlation is thus expected between DiP values and T$_\textrm{ME}$. Because of aging, in theory, the influence of ME expansion shall increase for a more extended ME. However, since we do not find such a correlation, this is consistent with previous studies that mention aging is not the only origin (possible presence of an intrinsic asymmetry as well) and therefore can not solely explain the observed asymmetries within MEs \citep[e.g.,][]{demoulin2008,demoulin2020,nieveschinchilla2022,regnault2023}.

\section{Summary and Conclusion}\label{sec:SC}

\justify
In this study, we perform a statistical investigation of 28 ICMEs measured in situ between 0.23--0.83~au by the PSP spacecraft from 2018 October to 2022 August. The overarching goal is to address the radial evolution of ICME structures in the inner heliosphere, using the widespread spatial distribution of PSP observations. We examine the radial evolution of four ICME aspects: (i) ME complexity, (ii) ICME magnetic field, (iii) expansion of sheath, and (iv) distortion of ME magnetic field. The main results can be summarized as follows:


\begin{enumerate}
    
    \item We find that ME configurations show evolution with heliocentric distance. The likelihood of observing coherent configurations (F- or F$_\textrm{r}$) is higher close to the Sun, whereas the likelihood of observing complex configurations (C$_\textrm{x}$ and E$_\textrm{j}$) increases farther out. In addition, weaker ICMEs seem to be more complex in terms of ME configurations. We also observe that the post-ICME magnetic field undergoes a continued relaxation period even after the ICME passage to go back to the pre-ICME levels. The level of impact in the post-ICME magnetic field depends on the heliocentric distance of the ICME encounter. Farther out from the Sun (\textgreater0.58~au), the post-ICME magnetic field is considerably stronger on average (${\sim}$44{\%}) compared to the pre-ICME value than encounters close (\textless 0.56~au) to the Sun (13{\%}). Overall, the observations confirm previous findings that the ME complexity is strongly coupled to the heliocentric distance \citep[e.g.,][]{riley2013,scolini2022}.

    \item From multi-linear robust regression, we find the average ME magnetic field strength to scale as $r^{(-1.21\pm0.44)}$. This power-law index hints at a less steep fall-off of the magnetic field with heliocentric distance, compared to previous results \citep[e.g.,][]{gulisano2010,winslow2015,davies2022}. However, considering the large uncertainties assigned to the power-law index, similar to \citet{good2019,davies2022}, the derived statistical relationship is comparable with previous studies. Still, it suggests that the expansion close to the Sun is not as strong as expected in the period studied. 
    
    \item For normalized magnetic fluctuations within ICME sheaths, we find no correlations between fluctuations and heliocentric distance (PCC=0.12) and sheath duration (PCC=-0.14). However, we report a significant positive correlation between fluctuations and the level of enhancement in the ME magnetic field compared to the upstream value (PCC=\textbf{0.76}). This is in agreement with \citet{salman2021} who highlighted that the ME properties are prominent factors behind the observed sheath variability near 1~au.
   
   \item We find no sheaths for ICMEs encountered by PSP within 0.38~au. This finding is different compared to the approximate sheath formation distances reported in \citet{salman2020a,temmer2022}, but hints at the important contribution of the ME magnetic pressure and the corresponding expansion rate closer to the Sun to the sheath formation. We also observe a weak positive correlation between the sheath duration and heliocentric distance (PCC=0.35). However, the observed weak correlation can also be a byproduct of a small sample size with considerable event-to-event variability.
   
   \item We find the DiP values to vary between 0.15--0.56. Even though we have more DiP values \textless0.5 (18) than DiP values \textgreater0.5 (10), half of the DiP values (50{\%}) are around 0.50$\pm0.05$. We notice that the distribution of DiP values has no radial dependence (PCC=-0.18). We also observe no correlation between DiP values and the level of enhancement in the ME magnetic field compared to the background (PCC=-0.25) and duration of the ME (PCC=-0.14). To examine the second correlation, we only consider DiP values (\textless0.5) that correspond to compression at the front of the ME (an apparent feature of expansion). Since we do not see the expected correlation with ME duration, this is an indication that aging or time evolution alone is not enough to characterize the distortion of the ME magnetic field. This was mentioned in previous studies as well \citep[e.g.,][]{demoulin2008,demoulin2020,nieveschinchilla2022,regnault2023}.     
\end{enumerate}

\justify
The PSP ICMEs analyzed in this study provide a unique opportunity to investigate the radial evolution of ICME structures with a more continuous spread of observations. Statistical analyses based on such observations allow us to derive overall statistical trends for ICME radial evolution in the inner heliosphere. However, assumptions of simple geometry (not taking into account ICME shapes and curvatures) and quantitative limitations (small data set) leave room for future in-depth exploration. In addition, the lack of continuous good-quality plasma data limits our ability to perform a comprehensive analysis of the nature of the ICMEs. With an extended timeline and inclusion of SolO encountered events, this will be possible in upcoming projects.  

\section*{acknowledgments}

\justify
The authors wish to thank the PSP mission team for providing the data. The authors acknowledge using NASA/Goddard Space Flight Center's Space Physics Data Facility's CDAWeb service (available at \url{https://cdaweb.gsfc.nasa.gov/index.html/}). T.~S. and T.~N.~-C. acknowledge support from the Parker Solar Probe and Solar Orbiter missions. T.~S. and L.~J. thank the support of NASA's STEREO mission and Heliophysics Guest Investigator (HGI; no.\ 80NSSC23K0447) program. N.~L. acknowledges support from NASA grants 80NSSC20K0700 and 80NSSC20K0431. F.~C. acknowledges the financial support by an appointment to the NASA Postdoctoral Program at NASA Goddard Space Flight Center, administered by Oak Ridge Associated Universities through a contract with NASA. E.~D. acknowledges funding by the European Union (ERC, HELIO4CAST, 101042188). Views and opinions expressed are however those of the author(s) only and do not necessarily reflect those of the European Union or the European Research Council Executive Agency. Neither the European Union nor the granting authority can be held responsible for them. We also thank the anonymous
reviewer for critically reading the paper and suggesting improvements.


\bibliography{Salman}{}
\bibliographystyle{aasjournal}

\end{document}